\documentclass[conference]{IEEEtran}
\IEEEoverridecommandlockouts
\usepackage{cite}
\usepackage{amsmath,amssymb,amsfonts}
\usepackage{algorithmic}
\usepackage{graphicx}
\usepackage{textcomp}
\usepackage{xcolor}
\usepackage{adjustbox}
\usepackage{enumitem}
\usepackage[font=small,skip=0pt]{caption}
\usepackage{pgfplots}
\usepackage{pgfplotstable}
\usepackage{tikz}
\usepackage{graphicx}
\newcommand{\lw}{1.5pt} 

\tikzset{mark options={line width=1pt,solid}}%

\definecolor{myred}{rgb}{0.95,0.00000,0}%
\definecolor{mygray}{rgb}{0.9,0.9,0.9}%
\definecolor{mygreen}{rgb}{0,0.5,0}%
\definecolor{myblue}{rgb}{0.4,0.42,0.8}%
\definecolor{myblack}{rgb}{0.2,0.2,0.2}%
\definecolor{mypurple}{rgb}{0.45,0,0.9}%
\definecolor{mywhite}{rgb}{1,1,1}%
\definecolor{mymagenta}{rgb}{0.95,0,0.95}%

\usepackage{cite,algorithm,algorithmic,amsmath,amssymb,amsthm,empheq,mhsetup}
\usepackage{subfigure,amsfonts,balance}
\usepackage{epstopdf}
\usepackage{enumitem}
\usepackage{setspace}
\usepackage[left=0.649in, right=0.68in, top=0.72in, bottom=1.049in,centering]{geometry}

\def\BibTeX{{\rm B\kern-.05em{\sc i\kern-.025em b}\kern-.08em
    T\kern-.1667em\lower.7ex\hbox{E}\kern-.125emX}}

\DeclareMathOperator{\EEE}{\mathbb{E}}

\DeclareMathOperator{\F}{\mathcal{F}}

\DeclareMathOperator{\K}{\mathcal{K}}

\DeclareMathOperator{\z}{\mathbf{z}}

\DeclareMathOperator{\Hh}{\mathbf{h}}
\DeclareMathOperator{\HH}{\mathbf{H}}

\DeclareMathOperator{\rrr}{\mathbf{r}}

\DeclareMathOperator{\G}{\mathbf{G}}
\DeclareMathOperator{\D}{\mathbf{D}}

\DeclareMathOperator{\LL}{\mathcal{L}}

\DeclareMathOperator{\CN}{\mathcal{CN}}

\DeclareMathOperator{\x}{\mathbf{x}}

\DeclareMathOperator{\s}{\mathbf{s}}

\DeclareMathOperator{\U}{\mathbf{U}}

\DeclareMathOperator{\uu}{\mathbf{u}}

\DeclareMathOperator{\g}{\mathbf{g}}

\DeclareMathOperator{\Z}{\mathbf{Z}}

\DeclareMathOperator{\ETA}{\boldsymbol{\eta}}

\DeclareMathOperator{\ZETA}{\boldsymbol{\zeta}}

\begin{document}
\bstctlcite{IEEEexample:BSTcontrol}
\title{Serving Federated Learning and Non-Federated Learning Users: A Massive MIMO Approach \vspace{-5mm}}
\author{\IEEEauthorblockN{Muhammad Farooq\IEEEauthorrefmark{1}, Tung T. Vu\IEEEauthorrefmark{2},
Hien Quoc Ngo\IEEEauthorrefmark{2}, Le-Nam Tran\IEEEauthorrefmark{1}}
\IEEEauthorblockA{{\small{}{}\IEEEauthorrefmark{1}School of Electrical and Electronic
Engineering, University College Dublin, Ireland}}
\IEEEauthorblockA{{\small{}{}\IEEEauthorrefmark{2}Institute of Electronics, Communications,
and Information Technology (ECIT), Queen's University Belfast, Belfast
BT3 9DT, UK}}  \IEEEauthorblockA{ Email: muhammad.farooq@ucdconnect.ie, t.vu@qub.ac.uk, hien.ngo@qub.ac.uk,
nam.tran@ucd.ie } \vspace{-10mm}}
\maketitle
\allowdisplaybreaks
\begin{abstract}
Federated learning (FL) with its data privacy protection and communication efficiency has been considered as a promising learning framework for beyond-5G/6G systems. We consider a scenario where a group of downlink non-FL users are jointly served with a group of FL users using massive multiple-input multiple-output technology. The main challenge is how to utilise the resource to optimally serve both FL and non-FL users. We propose a communication scheme that serves the downlink of the non-FL users (UEs) and the uplink of FL UEs in each half of the frequency band.  We formulate an optimization problem for optimizing transmit power to maximize the minimum effective data rates for non-FL users, while guaranteeing a quality-of-service time of each FL communication round for FL users. Then, a successive convex approximation-based algorithm is proposed to solve the formulated problem. Numerical results confirm that our proposed scheme significantly outperforms the baseline scheme. 
\end{abstract}

\vspace{-4mm}
\section{Introduction}
\vspace{-2mm}
\label{sec:Introd}
Federated learning (FL) has emerged as a promising machine learning framework in future wireless networks with a wide range of real-world digital applications such as Gboard, FedVision, etc. \cite{Aledhari20FL}. FL is an iterative process with many communication rounds. In each communication round, a central server sends a downlink global update to the users (UEs). The UEs compute their uplink local updates using their own data sets, and then send the local updates back to the central server for further updating the global learning model. The FL process continues until a specific learning accuracy level is achieved. By not sending the raw data, the data privacy is preserved and the communication traffic is significantly reduced. On the other hand, it is anticipated that  future wireless systems not only serve FL UEs but also non-FL UEs. In each FL communication round, FL UEs need both downlink and uplink, while non-FL UEs may only need one of these transmissions. This calls for a novel network design to support both groups of FL and non-FL UEs at the same time. 

In the related literature, previous publications studying the implementation of FL in wireless networks can be classified into learning-oriented or communication-oriented. The former aims to improve the learning performance (e.g., test accuracy) under the presence of detrimental factors of wireless systems such as thermal noise, fading, and estimation errors \cite{chen21TWC,amiri20TWC}. On the other hand, the latter focuses on enhancing the communication performance (e.g., execution time reduction, energy efficiency) \cite{vu20TWC,vu21ICC}. However, all above works consider the cases where only FL UEs are served.

\textit{Paper Contributions:} Motivated by communication-oriented studies, we propose a novel network design for serving FL and downlink non-FL UEs at the same time.
First, we leverage massive multiple-input multiple-output (MIMO) and assume that each FL communication round is executed in one large-scale coherence time.
Here, in the downlink of each FL communication round, both FL and non-FL groups are jointly served simultaneously and in the same frequency band. However, in the uplink of each FL communication round, the uplink transmission of FL UEs and the downlink transmission of non-FL UEs requires a two-way communication. Thus, we propose to serve these two transmissions separately at each half of the frequency band. Zero-forcing (ZF) processing is then used for both downlink and uplink transmissions. 
Next, we formulate an optimization problem that allocates power and computing resources to maximize the minimum effective data rate of non-FL users, while ensuring a quality-of-service execution time of each FL communication round for FL UEs. A successive convex approximation algorithm is then derived to solve the formulated problem. Numerical results show that our proposed scheme outperforms the considered baseline scheme. 

\vspace{-2mm}
\section{Proposed Scheme and System Model}
\vspace{-2mm}
\label{sec:SystModel} 

We consider a massive MIMO system \cite{sadeghi18TWC} where a single $M$-antenna BS serves simultaneously two groups of single-antenna non-FL and FL UEs. Further, the non-FL UEs are assumed to receive  data in the downlink.

Let $\LL\triangleq\{1,\dots,L\}$, and $\K\triangleq\{1,\dots,K\}$ be the sets of FL UEs and non-FL UEs, respectively. 
The FL framework of the FL group has the following four steps in each communication round \cite{tran19INFOCOM,vu20TWC,mcmahan17AISTATS}. 
\vspace{0mm}\begin{enumerate}[label=(S\arabic*)]
\item A central server sends a global update to FL UEs. 
\item The FL UEs compute their uplink local updates based on the global update and their local data. 
\item The uplink local updates are sent to the central server. 
\item The central server computes the global update using the received uplink local updates. 
\end{enumerate}\vspace{0mm}
Here, the BS acts as the central server. The non-FL UEs are assumed to keep receiving downlink data in every step of each FL communication round of FL UEs.

\vspace{-2mm}
\subsection{Proposed Scheme to Serve FL and downlink non-FL UEs}
\vspace{-2mm}
We assume that each FL communication round (instead of the
whole FL process) is executed in one large-scale coherence time. We then propose a synchronous scheme to support FL communication rounds. Here, all the FL UEs start each step of one FL communication round at the same time, and wait for others to start a new step. The global and local updates in Steps (S1) and (S3) can be transmitted in one or multiple (small-scale) coherence times based on their sizes. Each coherence time in Step (S1) or (S3) includes channel estimation and downlink or uplink transmission. 

The non-FL group is served along with the FL group as follows. In step (S1), the downlink of both  non-FL and FL groups are served at the same time and frequency band. In step (S2), the BS only serves non-FL UEs in the downlink. Then, in step (S3), owing to the half-duplex operation at the BS, the downlink of non-FL UEs and the uplink of FL UEs are served simultaneously but separately in each half of the frequency band.  

\vspace{-1mm}
\subsection{Detailed System Model of Proposed Scheme}
\vspace{-1mm}
\subsubsection{Step (S1)}
In this step, the BS sends the global update to FL UEs and the downlink data to non-FL UEs. To do this, the BS fist needs to acquire the channels via uplink training, and then uses these channel estimates to send the payload data (i.e., the global update or downlink data) to the UEs.

\textbf{Channel estimation}: For each coherence block of length $\tau_c$, the BS estimates the channels by using
uplink pilots received from all the UEs with a time-division-duplexing (TDD) protocol. Let $\sqrt{\rho_{p}}\boldsymbol{\varphi}_{\ell}\in\mathbb{C}^{\tau_{d,p}\times 1},\Vert\boldsymbol{\varphi}_{\ell}\Vert^{2}=\tau_{d,p}$
be the dedicated pilot symbols assigned to FL UE $\ell$,
and $\sqrt{\rho_{p}}\bar{\boldsymbol{\varphi}}_{k}\in\mathbb{C}^{\tau_{1,p}\times1},\Vert\bar{\boldsymbol{\varphi}}_{k}\Vert^{2}=\tau_{1,p}$
be the pilot sequence assigned to non-FL UE $k$, where
$\rho_{p}$ is the normalized transmit power of each pilot symbol, and $\tau_{d,p},\tau_{1,p}\geq L+K$ are the length of pilot sequences.
We assume the pilots of non-FL UEs and FL UEs are pairwisely orthogonal to avoid pilot contamination,
i.e. $\boldsymbol{\varphi}_{\ell}^{H}\bar{\boldsymbol{\varphi}}_{k}=0,\forall\ell,\forall k$,
$\boldsymbol{\varphi}_{\ell}^{H}\boldsymbol{\varphi}_{\ell^{\prime}}=0,\forall\ell^{\prime}\neq\ell$
and $\bar{\boldsymbol{\varphi}}_{k}^{H}\bar{\boldsymbol{\varphi}}_{k^{\prime}}=0,\forall k^{\prime}\neq k$.

Let $\G_{d}=[\g_{d,1},\dots,\g_{d,L}]\in\mathbb{C}^{M\times L}$ and
$\HH_{1}=[\Hh_{1,1},s,\Hh_{1,K}]\in\mathbb{C}^{M\times K}$ be
the channel matrices from the BS to the FL and non-FL groups in
Step (S1), respectively. Here, $\g_{d,\ell}=\beta_{\ell}^{1/2}{\z}_{d,\ell}$ represents the channel vector between the BS and FL UE $\ell$, while $\Hh_{k}=\bar{\beta}_{k}^{1/2}{\z}_{1,k}$ is the channel vector between the BS and non-FL UE $k$, where $\beta_{\ell}$, $\bar{\beta}_{k}$ are the large-scale fading coefficients, $\tilde{\z}_{d,\ell}$ and ${\z}_{1,k}$ are small-scale fading coefficients. Assuming minimum mean square error (MMSE) estimation, the channel estimate of
$\g_{d,\ell}$ can be written as $\check{\g}_{d,\ell}=\sigma_{d,\ell}\z_{d,\ell}$
where $\z_{d,\ell}\sim\mathcal{CN}(\mathbf{0},\mathbf{I}_{M})$ and $\sigma_{d,\ell}^{2}=\tfrac{\rho_{p}\tau_{d,p}\beta_{\ell}^{2}}{\rho_{p}\tau_{d,p}\beta_{\ell}+1}$ \cite{ngo16}. Similarly,
the estimate of $\Hh_{k}$ in Step (S1) can be
written as $\check{\Hh}_{1,k}=\sigma_{1,k}\z_{1,k}$, where $\z_{1,k}\sim\CN(\mathbf{0},\mathbf{I}_{M})$ and
 $\sigma_{1,k}^{2}=\tfrac{\rho_{p}\tau_{1,p}\bar{\beta}_{k}^{2}}{\rho_{p}\tau_{1,p}\bar{\beta}_{k}+1}$.
Let 
$\Z_{d}=[\z_{d,1},\dots,\z_{d,L}]$, $\Z_{1}=[\z_{1,1},\dots,\z_{1,K}]$.

\textbf{Downlink transmission for both FL and non-FL UEs}: The BS
encodes the global training update intended for FL UE $\ell$ into symbol
$s_{d,\ell}$, where $\EEE\{|s_{d,\ell}|^{2}\}\!=\!1$, $\forall\ell\!\in\!\mathcal{L}$, and encodes the
downlink data desired for non-FL UE $k$ into symbol
$s_{1,k}$, where $\EEE\{|s_{1,k}|^{2}\}\!=\!1$, $\forall k\!\in\!\K$. The data symbols are then precoded before being transmitted. Let $\s_{d}\triangleq[s_{d,1},\dots,s_{d,L}]^{T}$, $\s_{1}\triangleq[s_{1,1},\dots,s_{1,K}]^{T}$. $\eta_{d,\ell}$ and $\zeta_{1,k}$ denote the power control coefficient
associated with FL UE $\ell$ and non-FL UE $k$, respectively. Let $\U_{d}\triangleq[\uu_{d,1},\dots,\uu_{d,L}]$ and $\U_{1}\triangleq[\uu_{d,L+1},\dots,\uu_{d,L+K}]$ be precoding matrices for the two groups. Then, the transmitted signal at the BS in Step (S1) is given by $\x_{1}\!=\!\sqrt{\rho_{d}}\U_{d}\D_{\ETA_{d}}^{1/2}\s_{d}+\sqrt{\rho_{d}}\U_{1}\D_{\ZETA_{1}}^{1/2}\s_{1}$, where  $\ETA_{d}\triangleq [\eta_{d,1},\dots,\eta_{d,L}]^{T}$ and $\ZETA_{1}\triangleq[\zeta_{1,1},\dots,\zeta_{1,K}]^{T}$, and $\D_{\x}$ is the diagonal matrix with the elements of $\x$ on its
diagonal.
In this paper,  zero-forcing precoding
is applied to precode the symbols for all the FL and non-FL UEs.
Thus, the precoding vector $[\U_{d} ~ \U_{1}]\!=\!\sqrt{(M-L-K)}\Z(\Z^{H}\Z)^{-1}$
\cite[(3.49)]{ngo16}, where $\Z\!=\![\Z_{d},\Z_{1}]$, and $M\geq L+K$ is required.

The transmitted power at the BS is required to meet the average normalized
power constraint, i.e., $\EEE\{|\mathbf{x}_{1}|^{2}\}\leq\rho_{d}$,
which can be expressed as: 
\begin{equation}
\textstyle\sum\nolimits _{\ell\in\mathcal{L}}\eta_{d,\ell}+\textstyle\sum\nolimits _{k\in\K}\zeta_{1,k}\leq1.\label{powerdupperbound}
\end{equation}
Following \cite[Sec. 3.3.2]{ngo16}, the achievable rate for FL UE $\ell$ is given by $R_{d,\ell}(\ETA_{d},\ZETA_{1})\!=\! \tfrac{\tau_{c}-\tau_{d,p}}{\tau_{c}}B\log_{2}\!\big(1\!+\!\gamma_{d,\ell}(\ETA_{d},\ZETA_{1})\big)$,
where $B$ is the bandwidth, and $\gamma_{d,\ell}(\ETA_{d},\ZETA_{1})\!=\!
\tfrac{\rho_{d}\eta_{d,\ell}(M-L-K)\sigma_{d,\ell}^{2}}{1+\rho_{d}(\beta_{\ell}-\sigma_{d,\ell}^{2})\!\sum_{\ell\in\LL}\eta_{d,\ell}+\rho_{d}(\beta_{\ell}-\sigma_{d,\ell}^{2})\sum_{k\in\K}\zeta_{1,k}}$ is the effective signal-to-interference-plus-noise ratio (SINR).
Since all the FL UEs start and end a step synchronously, it is practically meaningful to define  the effective achievable rate (bps) of each FL UE to be the minimum achievable rate of the FL group, i.e., 
$\!R_{d}(\ETA_{d},\ZETA_{1})\!=  \min_{\ell\in\LL}R_{d,\ell}(\ETA_{d},\ZETA_{1}).$ Similarly, the achievable rate of non-FL UE $k$
is expressed as $R_{1,k}(\ETA_{d},\ZETA_{1})\!=\!\tfrac{\tau_{c}-\tau_{1,p}}{\tau_{c}}B\log_{2}\big(1\!+\!\gamma_{1,k}(\ETA_{d},\ZETA_{1})\big)$,
where the effective $\gamma_{1,k}(\ETA_{d},\ZETA_{1})$ is given by \cite[Sec. 3.3.2]{ngo16}: $\tfrac{\rho_{d}\zeta_{1,k}(M-L-K)\sigma_{1,k}^{2}}{1+\rho_{d}(\bar{\beta}_{k}-\sigma_{1,k}^{2})\sum_{k\in\K}\zeta_{1,k}+\rho_{d}(\bar{\beta}_{k}-\sigma_{1,k}^{2})\sum_{\ell\in\LL}\eta_{d,\ell}}$.
 
\textbf{Downlink delay of the FL group}: Let $S_{d}$ (bits) be the
data size of the global training update of the FL group. The transmission
time from the BS to FL UE $\ell\in\LL$ is given by $t_{d}(\ETA_{d},\ZETA_{1})=\tfrac{S_{d}}{R_{d}(\ETA_{d},\ZETA_{1})},\forall\ell$. 

\vspace{0.5mm}
\textbf{Amount of downlink data received at the non-FL UEs}:

The amount of data received at non-FL UE $k\in\K$ in step (S1) is $D_{1,k}(\ETA_{d},\ZETA_{1})=R_{1,k}(\ETA_{d},\ZETA_{1})t_{d}(\ETA_{d},\ZETA_{1})$.

\vspace{0mm}
\subsubsection{Step (S2)}
After receiving the global update, each FL UE $\ell$ computes its
local training update on its local dataset, while each non-FL UE $k$
keeps receiving data from the BS.

\textbf{Local computation}: After receiving the global update, each
FL UE executes $N_{c}$ local computing rounds over its data set to
compute its local update. Let $c_{\ell}$ (cycles/sample) be the number of processing cycles for FL UE $\ell$ to process one data sample \cite{vu20TWC}.
Denote by $D_{\ell}$ (samples) and $f_{\ell}$ (cycles/s) the size
of the local data set and the computing frequency of FL UE $\ell$,
respectively. To provide synchronization in this step, we
choose $f_{l}$ such that $f_{\ell}=\tfrac{D_{\ell}c_{\ell}f}{\bar{D}\bar{c}}$, where $\bar{D}=\max_{\ell\in\LL}D_{\ell}$,
$\bar{c}=\max_{\ell\in\LL}c_{\ell}$, and $f$ is a frequency control
coefficient. The computation time is thus the same for  all the FL UEs, denoted by $t_{C}(f)$, which is given by 
$t_{C}(f)\!=\!t_{C,\ell}(f)\!=\!\tfrac{N_{c}D_{\ell}c_{\ell}}{f_{\ell}}\!=\!\tfrac{N_{c}\bar{D}\bar{c}}{f},\forall\ell\in\LL$ \cite{vu20TWC,tran19INFOCOM}.

\textbf{Channel estimation for non-FL UEs}: The channel estimation for the non-FL UEs in step (S2) is performed similarly as shown in step (S1) (i.e., based on MMSE estimation). Consequently, the estimate of $\Hh_{2,k}$ (i.e., the channel between the BS and non-FL UE $k$) can be written as $\check{\Hh}_{2,k}=\sigma_{2,k}\z_{2,k}$,
where $\z_{2,k}\sim\mathcal{CN}(\mathbf{0},\mathbf{I}_{M})$  and $\sigma_{2,k}^{2}=\tfrac{\rho_{p}\tau_{2,p}\bar{\beta}_{k}^{2}}{\rho_{p}\tau_{2,p}\bar{\beta}_{k}+1}$, where $\tau_{2,p}\geq K$ is the length of pilot sequence. Let $\Z_{2}=[\z_{2,1},\dots,\z_{2,K}]$.

\textbf{Amount of downlink data received at the non-FL group}: 
Let $\ZETA_{2}\triangleq[\zeta_{2,1},\dots,\zeta_{2,K}]^{T}$ be the
power control coefficients for non-FL UEs in Step (S2). The average normalized power constraint at the BS can be expressed as
\vspace{-1mm}\begin{equation}
\textstyle\sum\nolimits _{k\in\K}\zeta_{2,k}\leq1.\label{powerdupperbound-1}
\end{equation}
Since there is no interference from the FL UEs to non-FL UEs in this step, the achievable downlink rate (bps) of non-FL UE $k,\forall k\in\K,$ is thus given by $R_{2,k}(\ZETA_{2}) =\tfrac{\tau_{c}-\tau_{2,p}}{\tau_{c}}B\log_{2}\big(1+\gamma_{2,k}(\ZETA_{2})\big)$,
where a ZF precoder vector $\uu_{2,k}\!=\!\sqrt{(M-K)} \Z_2(\Z_2^{H}\Z_2)^{-1} \mathbf{e}_{2,K}$ is applied for each non-FL UE $k\in\K$ at the BS, and $\gamma_{2,k}(\ZETA_{2})\!\!=\!\! \tfrac{\rho_{d}\zeta_{2,k}(M-K)\sigma_{2,k}^{2}}{1+\rho_{d}(\bar{\beta}_{k}-\sigma_{2,k}^{2})\sum_{k\in\K}\zeta_{2,k}}$. Thus, the amount of downlink data received at non-FL UE $k$ in step (S2) is $D_{2,k}(\ZETA_{2},f)=R_{2,k}(\ZETA_{2})t_{C}(f)$.

\subsubsection{Step (S3)}
In this step, the  local updates of the FL UEs are transmitted to the BS in the uplink, while downlink data are kept being sent from the BS to the non-FL UEs. 
The uplink transmission of the FL UEs and the downlink transmission of non-FL UEs are executed in two separate halves of the frequency band. 
\vspace{-1mm}

\textbf{Channel estimation}: Similar to the channel estimation in Steps (S1) and (S2), the channel $\g_{u,\ell}$ between the BS and the FL-UE
$\ell$ in Step (S3) has an estimate $\check{\g}_{u,\ell}\!=\!\sigma_{u,\ell}\z_{u,\ell}$,
where $\z_{u,\ell}\!\sim\!\mathcal{CN}(\mathbf{0},\mathbf{I}_{M})$  and $\sigma_{u,\ell}^{2}\!=\!\tfrac{\rho_{p}\tau_{u,p}\beta_{\ell}^{2}}{\rho_{p}\tau_{u,p}\beta_{\ell}+1}$. Here, $\tau_{u,p}\!\geq\! L+K$ is the length of pilot sequence.
The channel $\Hh_{3,k}$ between the BS and non FL-UE $k$
in Step (S3) has an estimate $\check{\Hh}_{3,k}\!=\!\sigma_{3,k}\z_{3,k}$,
where $\z_{3,k}\!\sim\!\mathcal{CN}(\mathbf{0},\mathbf{I}_{M})$  and $\sigma_{3,k}^{2}\!=\!\tfrac{\rho_{p}\tau_{3,k}\bar{\beta}_{k}^{2}}{\rho_{p}\tau_{3,k}\bar{\beta}_{k}+1}$.
Here, $\tau_{3,p}\!\geq \!L+K$ is the length of pilot sequence. Let $\Z_{u}\!\triangleq\![\z_{u,1},\dots,\z_{u,L}]$.

\textbf{Uplink transmission of FL UEs}: After computing the local
update, FL UE $\ell$ encodes this update into symbol
$s_{u,\ell}$, where $\EEE\{|s_{u,\ell}|^{2}\}=1$, and sends baseband signal  $x_{u,\ell}=\sqrt{\rho_{u}\eta_{u,\ell}}s_{u,\ell}$
to the BS, where $\eta_{u,\ell}$ is the power control coefficient and $\rho_{u}$ is the normalized uplink transmit power constraint, i.e., $\EEE\left\{ |x_{u,\ell}|^{2}\right\} \leq\rho_{u}$. Thus, we have the constraint: 
\vspace{-1mm}\begin{equation}
\eta_{u,\ell}\leq1,\forall\ell\in\LL.\label{poweruupperbound}
\end{equation}

Since the FL and non-FL groups are served in two separate halves of the frequency band, there is no interference among FL and non-FL groups. Thus, the achievable rate for FL UE $\ell$ is given by \cite[Sec. 3.3.2]{ngo16}: $R_{u,\ell}(\ETA_{u})\!=\! \tfrac{\tau_{c}-\tau_{u,p}}{\tau_{c}}\tfrac{B}{2}\log_{2}\!\big(1\!+\!\gamma_{u,\ell}(\ETA_{u})\big)$, 
where the ZF precoding vector applied for FL UE $\ell$ is  $\uu_{u,\ell}=\sqrt{(M-L)}\Z_{u}(\Z_{u}^{H}\Z_{u})^{-1}\mathbf{e}_{\ell,L}$ and $\gamma_{u,\ell}(\ETA_{u}) \!=\!\tfrac{\rho_{u}\eta_{u,\ell}(M-L)\sigma_{u,\ell}^{2}}{1+\rho_{u}\sum_{i\in\LL}(\beta_{i}-\sigma_{u,i})\eta_{u,i}}$. For synchronization, we choose
the rates of FL UEs to be the same as the minimum achievable rates
in the FL group, i.e., 
$ R_{u}(\ETA_{u})=\min_{\ell\in\LL}R_{u,\ell}(\ETA_{u}).
$

\textbf{Uplink delay of FL UEs}: Denote by $S_{u}$ (bits) the data size of
the local training update of the FL group. The transmission time from
FL UE $\ell$ to the BS is the same and given by $t_{u}(\ETA_{u})=\tfrac{S_{u}}{R_{u}(\ETA_{u})}$.

\textbf{Downlink transmission for Non-FL UEs}: Denote by $\zeta_{3,k}$
the power control coefficient of non-FL UE $k$ and $\ZETA_{3}\triangleq[\zeta_{3,1},\dots,\zeta_{3,K}]^{T}$. 
A ZF precoding matrix $\U_{3}=\sqrt{(M-K)}\Z_{3}(\Z_{3}^{H}\Z_{3})^{-1}$ is applied for non-FL UEs. The transmitted signal from the BS to the non-FL UEs is given as $\x_{3}=\sqrt{\rho_{d}}\U_{3}\D_{\ZETA_{3}}^{1/2}\s_{3}$, where $\s_{3}\triangleq[s_{3,1},\dots,s_{3,K}]^{T}$ and  $\EEE\{|s_{3,k}|^{2}\}=1$.
The transmitted power at the BS is constrained as $\EEE\{|\x_{3}|^{2}\}\leq\rho_{d}$, which
can be expressed as: 
\vspace{-1mm}\begin{equation}
\textstyle\sum\nolimits _{k\in\K}\zeta_{3,k}\leq1.\label{powerdupperbound-2}
\end{equation}
The achievable downlink rate for non-FL UE $k,\forall k\in\K,$
is $R_{3,k}(\ZETA_{3}) \!=\!\tfrac{\tau_{c}-\tau_{3,p}}{\tau_{c}}\tfrac{B}{2}\log_{2}\big(1\!+\!\gamma_{3,k}(\ZETA_{3})\big)$, 
where $\gamma_{3,k}(\ZETA_{3})\nonumber
 =\tfrac{\rho_{d}\zeta_{3,k}(M-K)\sigma_{3,k}^{2}}{1+\rho_{d}(\bar{\beta}_{k}-\sigma_{3,k}^2)\sum_{i\in\K}\zeta_{3,i}}$.

\textbf{Amount of downlink data received at the non-FL group}: The
amount of downlink data received at non-FL UE $k,\forall k\in\K$,
in Step (S3) is $D_{3,k}(\ETA_{u},\ZETA_{3})=R_{3,k}(\ZETA_{3})t_{u}(\ETA_{u})$.

\subsubsection{Step (S4)}

After receiving all the local update, the BS computes its global update. Since the computational capability of the central server is much more powerful than that of the UEs, the delay of computing the global update is negligible.

\vspace{-2mm}
\section{Problem Formulation and Solution}
\label{sec:PF}
\vspace{-1mm}
\subsection{Problem Formulation}
\vspace{-1mm}
In this section, we aim to maximize the minimum effective downlink rates of non-FL UEs while guaranteeing the quality of service (QoS) of the execution time of one FL communication round for FL UEs. Here, an effective downlink rate is defined as the ratio of the total received data of a non-FL UE and the corresponding transmission time (i.e., the total time of steps (S1)-(S3) of that FL UE). 
The considered problem can be formulated as  
\begin{subequations}
\label{Pmain} 
\begin{align}
\!\!\!\!\!\!\!\!\underset{\x}{\max}\,\, & \min_{k\in\K} \tfrac{D_{1,k}(\ETA_{d},\ZETA_{1})+D_{2,k}(f,\ZETA_{2})+D_{3,k}(\ETA_{u},\ZETA_{3})}{t_{d}(\ETA_{d},\ZETA_{1})+t_{C}(f)+t_{u}(\ETA_{u})}\\
\nonumber
\mathrm{s.t.}\,\, & \eqref{powerdupperbound},\eqref{powerdupperbound-1}, \eqref{poweruupperbound}, \eqref{powerdupperbound-2}
\\
\label{S1powerUbound}
 & \eta_{d,\ell}\geq0, \zeta_{1,k}\geq0, \zeta_{2,k}\geq0, \eta_{u,\ell}\geq0, \zeta_{3,k}\geq0\\
 & f_{\min}\leq f_{\ell}\leq f_{\max},\forall\ell\label{fbound}\\
 & t_{d}(\ETA_{d},\ZETA_{1})+t_{C}(f)+t_{u}(\ETA_{u})\leq t_{\text{QoS}},\label{eq:QoSbound}
\end{align}
\end{subequations}
where $\x\!\triangleq\!\!\{\ETA_{d},\!\ZETA_{1},\!f,\!\ZETA_{2},\!\ETA_{u},\!\ZETA_{3}\!\}$.  
Constraint \eqref{eq:QoSbound} ensures that the execution time of one FL communication round of each FL UE must not be greater than a pre-determined QoS threshold $t_{\text{QoS}}$. 
\vspace{-3mm}
\subsection{Solution}
\vspace{-1mm}
In the following, we propose a solution to \eqref{Pmain} based on
successive convex approximation (SCA) technique. First, we equivalently rewrite  problem \eqref{Pmain} into a more tractable  form as 
\vspace{-5mm}\begin{subequations}
\label{Pmain-2} 
\begin{align}
\!\!\!\underset{\bar{\x}}{\max}\,\, & t/t_{\text{Q}}
\\
\nonumber
\!\!\!\mathrm{s.t.}\,\, & 
\eqref{powerdupperbound},\eqref{powerdupperbound-1}, \eqref{poweruupperbound}, \eqref{powerdupperbound-2}, \eqref{S1powerUbound}, \eqref{fbound},\eqref{eq:QoSbound-2}
\\
&\!\tfrac{R_{1,k}(\ETA_{d},\ZETA_{1}\!)}{R_{d}(\ETA_{d},\ZETA_{1}\!)}\!S_{d}\!+\!R_{2,k}(\ZETA_{2})\!\tfrac{N_{c}\bar{D}\bar{c}}{f}\!+\!\tfrac{R_{3,k}(\ZETA_{3})}{R_{u}(\ETA_{u})}\!S_{u}\!\geq\!t,\!\forall k
\label{eq:data_const-2}
\\
&\tfrac{S_{d}}{R_{d}(\ETA_{d},\!\ZETA_{1})}+\tfrac{N_{c}\bar{D}\bar{c}}{f}+\tfrac{S_{u}}{R_{u}(\ETA_{u})}\leq t_{\text{Q}}.\label{eq:QoSbound-3}
\\
& t_{\text{Q}}\leq t_{\text{QoS}},\label{eq:QoSbound-2}
\end{align}
\end{subequations}
where $\bar{\x}=\{\x,t,t_{\text{Q}}\}$, and $t,t_{\text{Q}}$ are newly introduced additional variables.
We further rewrite \eqref{Pmain-2} as
\vspace{-2mm}\begin{subequations}
\label{Pmain-3-1} 
\begin{align}
\!\!\!\!\!\underset{\tilde{\x}}{\max}\,\, & z \\
\mathrm{s.t.}\,\, 
& \eqref{powerdupperbound},\eqref{powerdupperbound-1}, \eqref{poweruupperbound}, \eqref{powerdupperbound-2}, \eqref{S1powerUbound}, \eqref{fbound},\eqref{eq:QoSbound-2}\nonumber 
\\
\label{z}
& z t_{\text{Q}} \leq t
\\
 & a_{1}S_{d}+a_{2}N_{c}\bar{D}\bar{c}+a_{3}S_{u}\geq t,\forall k\label{CFPmain-lowerbound}
 \\
 & \tfrac{S_{d}}{r_{d}}+\tfrac{N_{c}\bar{D}\bar{c}}{f}+\tfrac{S_{u}}{r_{u}}\leq t_{\text{Q}}\label{eq:QoSbound-4}
 \\
 & r_{d}\leq R_{d,\ell}(\ETA_{d},\ZETA_{1});r_{u}\leq R_{u,\ell}(\ETA_{u}),\forall\ell\label{Rdell-lowerbound}\\
 & a_{1} \tilde{r}_{d} \leq r_{1,k};a_{2} f \leq r_{2,k};a_{3} \tilde{r}_{u} \leq r_{3,k},\forall k
 \label{ratioS1-lowerbound}
 \\
 & r_{1,k}\leq R_{1,k}(\ETA_{d},\ZETA_{1});r_{2,k}\leq R_{2,k}(\ZETA_{2});\nonumber\\&r_{3,k}\leq R_{3,k}(\ZETA_{3}),\forall k\label{R1k-lowerbound}\\
 & R_{d,\ell}(\ETA_{d},\ZETA_{1})\leq\tilde{r}_{d};R_{u,\ell}(\ETA_{u})\leq\tilde{r}_{u},\forall\ell\label{Rdell-upperbound}
\end{align}
\end{subequations}
 where $\tilde{\x}\triangleq\{\bar{\x},r_{d},r_{u},a_{1},a_{2},a_{3},\rrr_{1},\rrr_{2},\rrr_{3},\tilde{r}_{d},\tilde{r}_{u},z\}$,
$\rrr_{1}=\{r_{1,k}\}$, $\rrr_{2}=\{r_{2,k}\}$, $\rrr_{3}=\{r_{3,k}\}$, and $r_{d}$, $r_{u}$, $a_{1}$, $a_{2}$, $a_{3}$, $\rrr_{1}$, $\rrr_{2}$, $\rrr_{3}$, $\tilde{r}_{d}$, $\tilde{r}_{u}$, and $z$ are additional variables. Problem \eqref{Pmain-3-1} is difficult to solve due to nonconvex constraints \eqref{z}, \eqref{Rdell-lowerbound}-\eqref{Rdell-upperbound}. To deal with these constraints, we apply the SCA method, which is detailed next.

For constraints in \eqref{Rdell-lowerbound}, 
\eqref{R1k-lowerbound}, we first see that each of the rates $R_{d,\ell}(\ETA_{d},\ZETA_{1})$, $R_{u,\ell}(\ETA_{u})$,
$R_{1,k}(\ETA_{d},\ZETA_{1})$, $R_{2,k}(\ZETA_{2})$, and $R_{3,k}(\ZETA_{3})$ can be written as $c\log(1+\tfrac{x}{y})$, where $c$ is the prelog factor, and $x$ and $y$ are the numerator and denominator of SINR. The convex lower bounds are found by $\log\!\big(1+\tfrac{x}{y}\big)\geq  \log\big(1+\tfrac{x^{(\!n\!)}}{y^{(\!n\!)}}\big)+\tfrac{2x^{(\!n\!)}}{(x^{(\!n\!)}+y^{(\!n\!)})}-\tfrac{(x^{(\!n\!)})^{2}}{(x^{(\!n\!)}+y^{(\!n\!)})x}-\tfrac{x^{(\!n\!)}y}{(x^{(\!n\!)}+y^{(\!n\!)})y^{(\!n\!)}},$
where $x>0,y>0$ \cite[(76)]{long21TCOM}. Therefore, $R_{d,\ell}(\ETA_{d},\ZETA_{1})$, $R_{u,\ell}(\ETA_{u})$,
$R_{1,k}(\ETA_{d},\ZETA_{1})$, $R_{2,k}(\ZETA_{2})$, and $R_{3,k}(\ZETA_{3})$ in constraints \eqref{Rdell-lowerbound}, 
\eqref{R1k-lowerbound} have the concave lower bounds
$\tilde{R}_{d,\ell}(\ETA_{d},\ZETA_{1})$, $\tilde{R}_{u,\ell}(\ETA_{u})$,
$\tilde{R}_{1,k}(\ETA_{d},\ZETA_{1})$, $\tilde{R}_{2,k}(\ZETA_{2})$,
and $\tilde{R}_{3,k}(\ZETA_{3})$ which are provided in \eqref{Rd_tilde_apprx}--\eqref{R3k_tilde_approx}.
\begin{figure*}
\begin{align}
\tilde{R}_{d,\ell}(\ETA_{d},\ZETA_{1})& =\tfrac{\tau_{c}-\tau_{d,p}}{\tau_{c}\log2}B\Big[\log\big(1 + \tfrac{\psi_{d,\ell}^{(\!n\!)}}{\theta_{d,\ell}^{(\!n\!)}}\big) + \tfrac{2\psi_{d,\ell}^{(\!n\!)}}{\psi_{d,\ell}^{(\!n\!)} + \theta_{d,\ell}^{(\!n\!)}} - \tfrac{(\psi_{d,\ell}^{(\!n\!)})^{2}}{(\psi_{d,\ell}^{(\!n\!)}  +  \theta_{d,\ell}^{(\!n\!)})\psi_{d,\ell}}  -  \tfrac{\psi_{d,\ell}^{(\!n\!)}\theta_{d,\ell}}{(\psi_{d,\ell}^{(\!n\!)}  +  \theta_{d,\ell}^{(\!n\!)})\theta_{d,\ell}^{(\!n\!)}}\Big] \leq R_{d,\ell}(\ETA_{d},\ZETA_{1})
\label{Rd_tilde_apprx}\\
\tilde{R}_{u,\ell}(  \ETA_{u}  )
&     =  \tfrac{\tau_{c}  -  \tau_{u,p}}{\tau_{c}\log2}\tfrac{B}{2}  \Big[  \log\big(1    +  \tfrac{\psi_{u,\ell}^{(\!n\!)}}{\theta_{u,\ell}^{(\!n\!)}}\big)+\tfrac{2\psi_{u,\ell}^{(\!n\!)}}{\psi_{u,\ell}^{(\!n\!)}+\theta_{u,\ell}^{(\!n\!)}}-\tfrac{(\psi_{u,\ell}^{(\!n\!)})^{2}}{(\psi_{u,\ell}^{(\!n\!)}+\theta_{u,\ell}^{(\!n\!)})\psi_{u,\ell}}-\tfrac{\psi_{u,\ell}^{(\!n\!)}\theta_{u,\ell}}{(\psi_{u,\ell}^{(\!n\!)}+\theta_{u,\ell}^{(\!n\!)})\theta_{u,\ell}^{(\!n\!)}}\Big] \leq R_{u,\ell}(\ETA_{u})
\label{Ru_tilde_approx}\\
\tilde{R}_{1,k}(\ETA_{d},\ZETA_{1}) & =\tfrac{\tau_{c}-\tau_{1,p}}{\tau_{c}\log2}B\Big[\log\big(1+\tfrac{\psi_{1,k}^{(\!n\!)}}{\theta_{1,k}^{(\!n\!)}}\big)+\tfrac{2\psi_{1,k}^{(\!n\!)}}{\psi_{1,k}^{(\!n\!)}+\theta_{1,k}^{(\!n\!)}}-\tfrac{(\psi_{1,k}^{(\!n\!)})^{2}}{(\psi_{1,k}^{(\!n\!)}+\theta_{1,k}^{(\!n\!)})\psi_{1,k}}-\tfrac{\psi_{1,k}^{(\!n\!)}\theta_{1,k}}{(\psi_{1,k}^{(\!n\!)}+\theta_{1,k}^{(\!n\!)})\theta_{1,k}^{(\!n\!)}}\Big] \leq R_{1,k}(\ETA_{d},\ZETA_{1})
\label{R1k_tilde_approx}\\
\tilde{R}_{2,k}(\ZETA_{2}) & =\tfrac{\tau_{c}-\tau_{2,p}}{\tau_{c}\log2}B\Big[\log\big(1+\tfrac{\psi_{2,k}^{(\!n\!)}}{\theta_{2,k}^{(\!n\!)}}\big)+\tfrac{2\psi_{2,k}^{(\!n\!)}}{\psi_{2,k}^{(\!n\!)}+\theta_{2,k}^{(\!n\!)}}-\tfrac{(\psi_{2,k}^{(\!n\!)})^{2}}{(\psi_{2,k}^{(\!n\!)}+\theta_{2,k}^{(\!n\!)})\psi_{2,k}}-\tfrac{\psi_{2,k}^{(\!n\!)}\theta_{2,k}}{(\psi_{2,k}^{(\!n\!)}+\theta_{2,k}^{(\!n\!)})\theta_{2,k}^{(\!n\!)}}\Big] \leq R_{2,k}(\ZETA_{2})
\label{R2k_tilde_approx}\\
\tilde{R}_{3,k}(\ETA_{u},\ZETA_{3}) & =\tfrac{\tau_{c}-\tau_{3,p}}{\tau_{c}\log2}\tfrac{B}{2}\Big[\log\big(1+\tfrac{\psi_{3,k}^{(\!n\!)}}{\theta_{3,k}^{(\!n\!)}}\big)+\tfrac{2\psi_{3,k}^{(\!n\!)}}{\psi_{3,k}^{(\!n\!)}+\theta_{3,k}^{(\!n\!)}}-\tfrac{(\psi_{3,k}^{(\!n\!)})^{2}}{(\psi_{3,k}^{(\!n\!)}+\theta_{3,k}^{(\!n\!)})\psi_{3,k}}-\tfrac{\psi_{3,k}^{(\!n\!)}\theta_{3,k}}{(\psi_{3,k}^{(\!n\!)}+\theta_{3,k}^{(\!n\!)})\theta_{3,k}^{(\!n\!)}}\Big] \leq R_{3,k}(\ZETA_{3})
\label{R3k_tilde_approx}
\end{align}
\vspace{-5mm}
\end{figure*}
Here $\psi_{d,\ell}=\!\rho_{d}(M\!-\!L\!-\!K)\sigma_{d,\ell}^{2}\eta_{d,\ell},
\psi_{d,\ell}^{(\!n\!)}\!=\!\rho_{d}(M\!-\!L\!-\!K)\sigma_{d,\ell}^{2}\eta_{d,\ell}^{(\!n\!)},
\theta_{d,\ell}\!=\!1\!+\!\rho_{d}(\beta_{\ell}\!-\!\sigma_{d,\ell}^{2})\sum_{i\in\LL}\eta_{d,i}\!+\!\rho_{d}\beta_{\ell}\sum_{k\in\K}\zeta_{1,k},
\theta_{d,\ell}^{(\!n\!)}\!=\!1\!+\!\rho_{d}(\beta_{\ell}\!-\!\sigma_{d,\ell}^{2})\sum_{i\in\LL}\eta_{d,i}^{(\!n\!)}\!+\!\rho_{d}\beta_{\ell}\sum_{k\in\K}\zeta_{1,k}^{(\!n\!)},
\psi_{u,\ell}\!=\!\rho_{u}(M\!-\!L)\sigma_{u,\ell}^{2}\eta_{u,\ell},
\psi_{u,\ell}^{(\!n\!)}\!=\!\rho_{u}(M\!-\!L)\sigma_{u,\ell}^{2}\eta_{u,\ell}^{(\!n\!)},
\theta_{u,\ell}\!=\!1\!+\!\rho_{u}\sum_{i\in\LL}(\beta_{i}\!-\!\sigma_{u,i}^{2})\eta_{u,i},
\theta_{u,\ell}^{(\!n\!)}\!=\!1\!+\!\rho_{u}\sum_{i\in\LL}(\beta_{i}\!-\!\sigma_{u,i}^{2})\eta_{u,i}^{(\!n\!)},
\psi_{1,k}=\rho_{d}(M\!-\!L\!-\!K)\sigma_{1,k}^{2}\zeta_{1,k},
\psi_{1,k}^{(\!n\!)}=\rho_{d}(M\!-\!L\!-\!K)\sigma_{1,k}^{2}\zeta_{1,k}^{(\!n\!)},
\theta_{1,k}=1\!+\!\rho_{d}(\bar{\beta}_{k}\!-\!\sigma_{1,k}^{2})\sum_{i\in\K}\zeta_{1,i}\!+\!\rho_{d}\bar{\beta}_{k}\sum_{\ell\in\LL}\eta_{d,\ell},
\theta_{1,k}^{(\!n\!)}=1\!+\!\rho_{d}
(\bar{\beta}_{k}\!-\!\sigma_{1,k}^{2})\sum_{i\in\K}\zeta_{1,i}^{(\!n\!)}\!+\!\rho_{d}\bar{\beta}_{k}\sum_{\ell\in\LL}\eta_{d,\ell}^{(\!n\!)};\psi_{2,k}=\rho_{d}(M\!-\! K)\sigma_{1,k}^{2}\zeta_{2,k},\psi_{2,k}^{(\!n\!)}=\rho_{d}(M\!-\!K)\sigma_{k}^{2}\zeta_{2,k}^{(\!n\!)},
\theta_{2,k}=1\!+\!\rho_{d}(\bar{\beta}_{k}\!-\!\sigma_{2,k}^{2})\sum_{i\in\K}\zeta_{2,i},
\theta_{2,k}^{(\!n\!)}=1\!+\!\rho_{d}(\bar{\beta}_{k}\!-\!\sigma_{2,k}^{2})\sum_{i\in\K}\zeta_{2,i}^{(\!n\!)},
\psi_{3,k}=\rho_{d}(M\!-\!K)\sigma_{3,k}^{2}\zeta_{3,k},
\psi_{3,k}^{(\!n\!)}=\rho_{d}(M\!-\!K)\sigma_{3,k}^{2}\zeta_{3,k}^{(\!n\!)},
\theta_{3,k}=1\!+\!\rho_{d}(\bar{\beta}_{k}\!-\!\sigma_{3,k}^{2})\sum_{i\in\K}\zeta_{3,i},
\theta_{3,k}^{(\!n\!)}\!=\!1\!+\!\rho_{d}(\bar{\beta}_{k}\!-\!\sigma_{3,k}^{2}) \sum_{i\in\K}\zeta_{3,i}^{(\!n\!)}$.
Therefore, constraints in \eqref{Rdell-lowerbound}, 
\eqref{R1k-lowerbound} can be approximated
by the following convex constraints 
\begin{subequations}
\label{eqs:LBs}
\begin{align}
r_{d}\!\leq & \tilde{R}_{d,\ell}(\ETA_{d},\ZETA_{1});
r_{u}\!\leq \! \tilde{R}_{u,\ell}(\ETA_{u}),\forall\ell;\\
\!\!r_{1,k}\!\leq & \tilde{R}_{1,k}(\ETA_{d},\!\ZETA_{1}\!);
r_{2,k}\!\leq \! \tilde{R}_{2,k}(\ZETA_{2});
r_{3,k}\!\leq \! \tilde{R}_{3,k}(\ZETA_{3}),\!\forall k.
\end{align}
\end{subequations}

For constraints \eqref{z}, \eqref{ratioS1-lowerbound}, we observe that $xy \leq \tfrac{1}{4} [(x+y)^2-2(x^{(\!n\!)}-y^{(\!n\!)})(x-y)
+ (x^{(\!n\!)}-y^{(\!n\!)})^2 ]$, where $x\geq0, y\geq0$ \cite{vu20TWC}. Therefore, \eqref{z}, \eqref{ratioS1-lowerbound} can be approximated by the following convex constraints
\begin{align}
\!&\!\!\!\tfrac{1}{4}[(z\!+\!t_{\text{Q}})^{2}\!\!-\!2(z^{(\!n\!)}\!-\!t_{\text{Q}}^{(\!n\!)})(z\!-\!t_{\text{Q}})\!+\!(z^{(\!n\!)}\!-\!t_{\text{Q}}^{(\!n\!)})^{2}]\!-\!t\!\leq \!0,
\label{z:approx}
\\
\!&\!\!\!\tfrac{1}{4}[(a_{1}\!\!+\!\tilde{r}_{d}\!)^{2} \!\!-\!2 (a_{1}^{(\!n\!)}\!\!-\!\tilde{r}_{d}^{(\!n\!)}\!)  (a_{1}\!\!-\!\tilde{r}_{d}\!)\!+\! (a_{1}^{(\!n\!)}\!\!-\!\tilde{r}_{d}^{(\!n\!)}\!)^{2}]\!\!-\!r_{1,k}\!\!\leq\! 0,
\label{ratioS1-lowerbound:approx}
\\
\!&\!\!\! \tfrac{1}{4}[(a_{2}\!+\!\!f)^{2}\!\!\!-\!2(a_{2}^{(\!n\!)}\!\!-\!f^{(\!n\!)}\!)(a_{2}\!-\!\!f) \!+\!(a_{2}^{(\!n\!)}\!\!-\!\!f^{(\!n\!)}\!)^{2}]\!\!-\! r_{2,k} \!\leq\! 0,
\\
\!&\!\!\! \tfrac{1}{4}[(a_{3}\!\!+\!\tilde{r}_{u}\!)^{2} \!\!-\! 2(a_{3}^{(\!n\!)}\!\!-\!\tilde{r}_{u}^{(\!n\!)}\!) (a_{3}\!\!-\!\tilde{r}_{u}\!) \!+\! (a_{3}^{(\!n\!)}\!\!-\!\tilde{r}_{u}^{(\!n\!)}\!)^{2}]\!\!-\! r_{3,k}\!\!\leq \!0.
\label{ratioS3-lowerbound:approx}
\end{align}

For constraints \eqref{Rdell-upperbound},
it is true that $\log\big(1+\tfrac{x}{y}\big)\leq  \log\big(1+\tfrac{x^{(\!n\!)}}{y^{(\!n\!)}}\big)+\tfrac{y^{(\!n\!)}}{(x^{(\!n\!)}+y^{(\!n\!)})}
\big(\tfrac{(x^{2}+(x^{(\!n\!)})^{2})}{2x^{(\!n\!)}y}-\tfrac{x^{(\!n\!)}}{y^{(\!n\!)}}\big)$,
where $x>0,y>0$
\cite[(75)]{sheng18TWC}.
Therefore, ${R}_{d,\ell}(\ETA_{u})$ and $R_{u,\ell}(\ETA_{u})$ in \eqref{Rdell-upperbound}  have the following convex upper bounds 
$\hat{R}_{d,\ell}(\ETA_{d},\ZETA_{1})$ and $\hat{R}_{u,\ell}(\ETA_{u})$, which
are given by \eqref{Rd_hat_approx} and \eqref{Ru_hat_approx}, respectively (see the top of the next page). 
\begin{figure*}
\begin{align}
R_{d}(\ETA_{d},\ZETA_{1}) \leq \hat{R}_{d}(\ETA_{d},\ZETA_{1}) & =\tfrac{\tau_{c}-\tau_{d,p}}{\tau_{c}\log2}B\Big[\log\big(1+\tfrac{\psi_{d,\ell}^{(\!n\!)}}{\theta_{d,\ell}^{(\!n\!)}}\big)+\tfrac{\theta_{d,\ell}^{(\!n\!)}}{\psi_{d,\ell}^{(\!n\!)}+\theta_{d,\ell}^{(\!n\!)}}-\tfrac{(\psi_{d,\ell})^{2}+(\psi_{d,\ell}^{(\!n\!)})^{2}}{2\psi_{d,\ell}^{(\!n\!)}\theta_{d,\ell}}-\tfrac{\psi_{d,\ell}^{(\!n\!)}}{\theta_{d,\ell}^{(\!n\!)}}\Big] 
\label{Rd_hat_approx}
\end{align}
\begin{align}
R_{u,\ell}(\ETA_{u}) \leq \hat{R}_{u}(\ETA_{u}) & =\tfrac{\tau_{c}-\tau_{u,p}}{\tau_{c}\log2}\tfrac{B}{2}\Big[\log\big(1+\tfrac{\psi_{u,\ell}^{(\!n\!)}}{\theta_{u,\ell}^{(\!n\!)}}\big)+\tfrac{\theta_{u,\ell}^{(\!n\!)}}{\psi_{u,\ell}^{(\!n\!)}+\theta_{u,\ell}^{(\!n\!)}}-\tfrac{(\psi_{u,\ell})^{2}+(\psi_{u,\ell}^{(\!n\!)})^{2}}{2\psi_{u,\ell}^{(\!n\!)}\theta_{u,\ell}}-\tfrac{\psi_{u,\ell}^{(\!n\!)}}{\theta_{u,\ell}^{(\!n\!)}}\Big]
\label{Ru_hat_approx}
\end{align}
\vspace{-2mm}
\hrulefill
\vspace{-2mm}
\end{figure*}
Therefore, constraints \eqref{Rdell-upperbound} 
can be approximated by the following convex constraints 
\begin{align}\label{eqs:UBs}
\hat{R}_{d,\ell}(\ETA_{d},\ZETA_{1}) & \leq\tilde{r}_{d};
\hat{R}_{u,\ell}(\ETA_{u})  \leq\tilde{r}_{u},\forall\ell.
\end{align}

At iteration $n+1$, for a given point $\tilde{\x}^{(0)}$,  problem \eqref{Pmain-2} is approximated by the following convex problem:
\begin{align}
{\max}\,\, & \{z \, |\, \tilde{\x}\in\tilde{\F}\},\label{eq:convexprob}
\end{align}
where $\tilde{\F}\triangleq \eqref{powerdupperbound},\eqref{powerdupperbound-1}, \eqref{poweruupperbound}, \eqref{powerdupperbound-2}, \eqref{S1powerUbound}, \eqref{fbound},\eqref{eq:QoSbound-2}, \eqref{CFPmain-lowerbound}, \eqref{eq:QoSbound-4}, \eqref{eqs:LBs}-\eqref{eqs:UBs}\}$. In Algorithm~\ref{alg:SCA}, we outline the main steps to solve problem \eqref{Pmain-3-1}. Here, $\F\triangleq\{\eqref{powerdupperbound},\eqref{powerdupperbound-1}, \eqref{poweruupperbound}, \eqref{powerdupperbound-2}, \eqref{S1powerUbound}, \eqref{fbound},\eqref{eq:QoSbound-2},\eqref{z}-\eqref{Rdell-upperbound}\}$ is the feasible set of problem \eqref{Pmain-3-1}.
In the case when $\tilde{\F}$ satisfies some constraint qualifications such as Slater's condition, Algorithm~\ref{alg:SCA} will converge to a stationary solution to \eqref{Pmain-3-1}.

\begin{algorithm}[t!]
\caption{Algorithm for solving \eqref{Pmain}}
\label{alg:SCA}
\begin{algorithmic}[1]
\STATE Input: Set $n=0$ and choose initial point $\tilde{\x}^{(0)}\in\F$
\REPEAT
\STATE Set $n=n+1$
\STATE Solve \eqref{eq:convexprob} to get $\tilde{\x}^\ast$
\STATE Set $\tilde{\x}^{(\!n\!)}=\tilde{\x}^\ast$
\UNTIL{convergence}
\end{algorithmic}
\end{algorithm}

\vspace{-5mm}
\section{Numerical Examples}
\vspace{-2mm}
\subsubsection{Parameter Setting}
\vspace{0mm}
We consider a $D\times D~\text{m}^{2}$ area where the BS is at the centre, while $L$ FL UEs and $K$ non-FL
UEs are randomly distributed. The large-scale fading coefficients are modeled in the same manner as \cite[Eq. (46)]{ziya21TCOM}: $\beta_k[\text{dB}] = - 148.1  - 37.6 \log_{10}\big(\tfrac{d_k}{1\,\,\text{km}}\big) + z_k$, 
where $d_k \geq 35$ m is the distance between UE $k$ and the BS, $z_k$ is a shadow fading coefficient which is modeled using a log-normal distribution having zero mean and $7$ dB standard deviation. We set $N_{0}=-92$ dBm, 
$t_{\text{QoS}}=3$ s, $B=20$ MHz, $\rho_{d}=10$ W, $\rho_{u}=\rho_{p}=0.2$
W, $\tau_{d,p}=\tau_{u,p}=20$, $\tau_{S_{1},p} =\tau_{S_{2},p} =\tau_{S_{3},p}=20$, $\tau_{c}=200$, $f_{\min}=0$, $f_{\max}=5\times10^{9}$ cycles/s, $D_{\ell}=D_{\max}=1.6\times10^5$ samples, $c_{\ell}=c_{\max}=20$ cycles/sample, $N_c=20$, $S_d=S_u=16\times10^6$ bits or 16Mb.

\vspace{0mm}
\subsubsection{Results and Discussions}
\vspace{0mm}
Since there are no other existing works that study massive MIMO networks for supporting both FL and non-FL groups, we compare our proposed scheme with a baseline scheme denotes as
\textbf{BL}. In \textbf{BL}, equal power allocation is adopted for both FL and non-FL UEs in step (S1), i.e., $\eta_{d,\ell}\!=\zeta_{1,k}=\!\tfrac{1}{L+K}, \forall \ell,k$. The same is applied to non-FL UEs in step (S2) and (S3), i.e, $\zeta_{2,k}=\zeta_{3,k}=\tfrac{1}{K}, \forall k$. In addition, in step (S3), each FL UE uses full power, i.e, $\eta_{u,\ell}=1, \forall \ell$.
    The processing frequencies are $f = \tfrac{N_c \bar{D} \bar{c}}{t_{\text{QoS}}-t_{d} - t_{u}}$. 

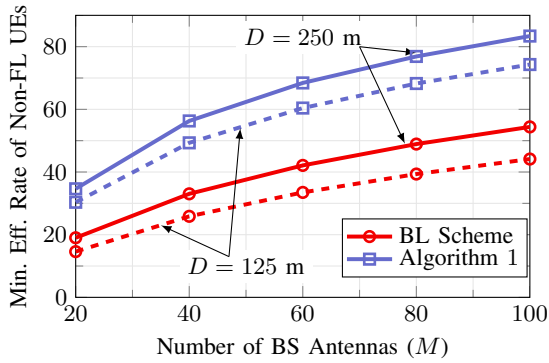
\begin{figure}[t!]
\vspace{-3mm}
	\centering
	\pgfplotsset{width=3in,height =0.7*3in, compat=1.6}
	\setlength{\baselineskip}{0pt}
\begin{tikzpicture}
{\small\begin{axis}[
xtick={20,40,60,80,100},
ymin=0,ymax=90,xmin=20,xmax=100,
grid=both,
minor tick num = 1,
grid style={mygray},
legend style={anchor=north east,draw=black,fill=white,legend cell align=left,inner sep=1pt,row sep = -3pt,at={(0.995,0.28)}},
xlabel={ Number of BS Antennas ($M$)},
ylabel={ Min. Eff. Rate of Non-FL UEs},
legend entries={BL Scheme, Algorithm $1$}
]
\addlegendimage{mark=o,line width=\lw, draw=myred}
\addlegendimage{mark=square,line width=\lw, draw=myblue}

\addplot[dashed,mark=o, line width=\lw, draw=myred] table [y=EPA11, x=iter,col sep = comma]{antennaJ4.csv};
\addplot[dashed,mark=square, line width=\lw, draw=myblue] table [y=HD11, x=iter,col sep = comma]{antennaJ4.csv};
\addplot[mark=o, line width=\lw, draw=myred] table [y=EPA22, x=iter,col sep = comma]{antennaJ4.csv};
\addplot[mark=square, line width=\lw, draw=myblue] table [y=HD22, x=iter,col sep = comma]{antennaJ4.csv};

\node[align=center,fill=white,inner sep=3pt] (NoPC) at (axis cs: 60,83) { $D=250$ m};
\draw[->,>=latex] (axis cs:70,80) -- (axis cs:80,78);
\draw[->,>=latex] (axis cs:70,80) -- (axis cs:78,49);

\node[align=center,fill=white,inner sep=3pt] (NoPC) at (axis cs: 50,10) { $D=125$ m};
\draw[->,>=latex] (axis cs:47,13) -- (axis cs:35,23);
\draw[->,>=latex] (axis cs:47,13) -- (axis cs:49,54);


\end{axis}}

\end{tikzpicture} 
	\caption{Minimum effective rate of non-FL UEs (Mbps) for different values of number of BS antennas. Here $L=K=5$.}
	\label{dataVSantennas} 
	\vspace{-3mm}
\end{figure}

\begin{figure}[t!]
\vspace{-3mm}
	\centering 
	\pgfplotsset{width=3in,height =0.7*3in, compat=1.6}
	\setlength{\baselineskip}{0pt}
\begin{tikzpicture}
{\small\begin{axis}[
xtick={2,4,...,8},
ymin=20,ymax=105,xmin=2,xmax=8,
grid=both,
minor tick num = 1,
grid style={mygray},
legend style={anchor=north east,draw=black,fill=white,legend cell align=left,inner sep=1pt,row sep = -3pt,at={(0.995,0.995)}},
xlabel={ Number of FL UEs ($L$)},
ylabel={ Min. Eff. Rate of Non-FL UEs},
legend entries={ BL Scheme, Algorithm $1$}
]
\addlegendimage{mark=o,line width=\lw, draw=myred}
\addlegendimage{mark=square,line width=\lw, draw=myblue}

\addplot[mark=o,dashed,line width=\lw, draw=myred] table [y=EPA1, x=L,col sep = comma]{FLUEsJ3.csv};
\addplot[mark=square,dashed,line width=\lw, draw=myblue] table [y=HD1, x=L,col sep = comma]{FLUEsJ3.csv};
\addplot[mark=o,line width=\lw, draw=myred] table [y=EPA2, x=L,col sep = comma]{FLUEsJ3.csv};
\addplot[mark=square,line width=\lw, draw=myblue] table [y=HD2, x=L,col sep = comma]{FLUEsJ3.csv};

%
%

\end{axis}

\begin{axis}[
ymin=20,ymax=105,xmin=2,xmax=8,
legend style={anchor=north east,draw=black,fill=white,legend cell align=left,inner sep=1pt,row sep = -3pt,at={(0.36,0.21)}},
legend entries={ $M=50$,  $M=100$}
]
\addlegendimage{dashed,line width=\lw, draw=myblack}
\addlegendimage{line width=\lw, draw=myblack}
\end{axis}}

\end{tikzpicture} 
	\caption{Minimum effective rate of non-FL UEs (Mbps) for different values of number of FL UEs. Here $K=5$, and $D=250$ m.}
	\label{datavsFLUEs} 
\end{figure}
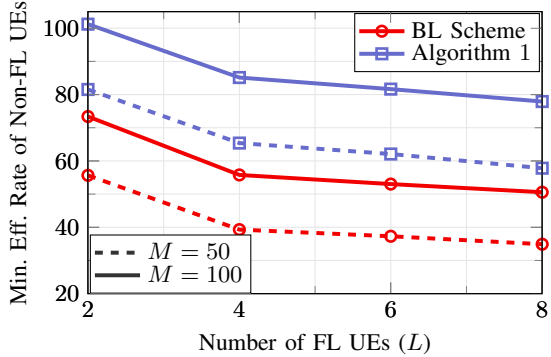

In Figs.~\ref{dataVSantennas} and~\ref{datavsFLUEs}, we compare the minimum effective rate of the non-FL UEs (in Mbps)  achieved by the proposed scheme and \textbf{BL}. As seen, the proposed scheme offers  better performance than the baseline scheme. The figures not only demonstrate a significant advantage of a joint allocation of power and computing frequency over  \textbf{BL}, which is heuristic, but also show the benefit of using massive MIMO. Specifically, thanks to massive MIMO technology, the data rate of each non-FL UE increases when the number of antennas increases, which then leads to a significant increase in the minimum effective data rates. We note that our proposed scheme also outperforms another baseline scheme using the frequency division multiple access (FDMA) approach in Step (S3). This baseline even provides a worse performance than that of the considered \textbf{BL} scheme, and hence, is skipped in this paper due to space limitation.
\vspace{-3mm}
\section{Conclusion}
\vspace{-2mm}
\label{sec:con} 
We have proposed a communication scheme using massive MIMO networks to serve FL and non-FL groups. Based on the successive convex approximation technique, we have  developed an algorithm to allocate transmit power and computing frequency in order to maximize the minimum effective rate of non-FL UEs while guaranteeing a quality-of-service execution time for each FL communication round of FL UEs. Numerical results have showed that the proposed scheme significantly improves the minimum effective rate of non-FL UEs compared to a baseline heuristic scheme.   
\vspace{-2mm}
\section*{Acknowledgment}
\vspace{-1mm}
The work of M. Farooq and L. N. Tran has emanated from research supported by a Grant from Science Foundation Ireland under Grant number 17/CDA/4786. The work of T. T. Vu and H. Q. Ngo was supported by the U.K. Research and Innovation Future Leaders Fellowships under Grant MR/S017666/1. 

\begin{spacing}{0.95}
\bibliographystyle{IEEEtran}
\bibliography{IEEEabrv,newidea2021}
\vspace{0mm}
\end{spacing}
\end{document}